# 2D and 3D CT Radiomic Features Performance Comparison in Characterization of Gastric Cancer: A Multi-center Study

Lingwei Meng, Di Dong, Xin Chen, Mengjie Fang, Rongpin Wang, Jing Li, Zaiyi Liu, and Jie Tian, *Fellow, IEEE*

*Abstract—* *Objective:* Radiomics, an emerging tool for medical image analysis, is potential towards precisely characterizing gastric cancer (GC). Whether using one-slice 2D annotation or whole-volume 3D annotation remains a long-time debate, especially for heterogeneous GC. We comprehensively compared 2D and 3D radiomic features' representation and discrimination capacity regarding GC, via three tasks ($T^{LNM}$, lymph node metastasis' prediction; $T^{LVI}$, lymphovascular invasion's prediction; $T^{pT}$, pT4 or other pT stages' classification). *Methods:* Four-center 539 GC patients were retrospectively enrolled and divided into the training and validation cohorts. From 2D or 3D regions of interest (ROIs) annotated by radiologists, radiomic features were extracted respectively. Feature selection and model construction procedures were customed for each combination of two modalities (2D or 3D) and three tasks. Subsequently, six machine learning models ($Model_{2D}^{LNM}$, $Model_{2D}^{LNM}$; $Model_{2D}^{LVI}$, $Model_{3D}^{LVI}$; $Model_{2D}^{pT}$, $Model_{3D}^{pT}$) were derived and evaluated to reflect modalities' performances in characterizing GC. Furthermore, we performed an auxiliary experiment to assess modalities' performances when resampling spacing different. *Results:* Regarding three tasks, the yielded areas under the curve (AUCs) were: $Model_{2D}^{LNM}$'s 0.712 (95% confidence interval, 0.613-0.811), $Model_{3D}^{LNM}$'s 0.680 (0.584-0.775); $Model_{2D}^{LVI}$'s 0.677 (0.595-0.761), $Model_{3D}^{LVI}$'s 0.615 (0.528-0.703); $Model_{2D}^{pT}$'s 0.840 (0.779-0.901), $Model_{3D}^{pT}$'s 0.813 (0.747-0.879). Moreover, the auxiliary experiment indicated that $Models_{2D}$ are statistically advantageous than $Models_{3D}$ with different resampling spacings. *Conclusion:* Models constructed with 2D radiomic features revealed comparable performances with those constructed with 3D features in characterizing GC. *Significance:* Our work indicated that time-saving 2D annotation would be the better choice in GC, and provided a related reference to further radiomics-based researches.

*Index Terms—* Radiomics features, gastric cancer, computed tomography (CT), computer-aided detection (CAD).

## I. Introduction

Gastric cancer (GC) is one of the most common malignant neoplasms and the third leading cause of cancer-related deaths worldwide [1]. The GC patients' prognosis is closely associated with the depth of tumor invasion, extent of lymph node metastasis (LNM) and presence of lymphovascular invasion (LVI) [2], [3]. Accurate preoperative evaluation of these prognostic factors is vital for GC patients in selecting appropriate treatment plans [4]. Computed tomography (CT) is the most common examination used in preoperative assessment due to stable image quality, high spatial resolution and fast acquisition speed [5], [6]. However, CT evaluation of the depth of tumor invasion and lymph nodes mainly depends on the size, morphology and enhancement pattern, which only have a moderate accuracy.

Recently, radiomics, the methodology of extracting a large panel of quantitative features from the conversions of imaging data and further data analysis for clinical decision support, increasingly attracts attention [7]–[9]. Radiomics enables the non-invasive profiling of tumor heterogeneity by combining multiple features in parallel. Many radiomics-based studies have provided some insights in oncologic practice related to cancer differential diagnosis, LNM, survival and therapeutic response evaluation [10]–[16]. *Wang et al.* [17] explored the CT radiomics approach's potential performance in the prediction of the depth of tumor invasion in GC. *Ma et al.* [18] found that the radiomic features could be used to differentiate GC from primary gastric lymphoma. *Jiang et al.* [19], [20] demonstrated that radiomic features could predict LNM, survival, and chemotherapeutic benefit in GC.

Radiomics comprises several steps, including tumor annotation, feature extraction, feature selection, and data mining [12]. Of the steps, selecting more reproducible and effective features is crucial for further data mining. However,

This work was supported by National Natural Science Foundation of China (91959130, 81971776, 81771924, 81601469, 81771912, 81501616, 81227901, 81671851, 81527805), Beijing Natural Science Foundation (L182061), National Science Fund for Distinguished Young Scholars (81925023), National Key R&D Program of China (2017YFC1308700, 2017YFA0205200, 2017YFC1309100, 2017YFA0700401), Science and Technology Planning Project of Guangzhou (201804010032), Bureau of International Cooperation of Chinese Academy of Sciences (173211KYSB20160053), Instrument Developing Project of Chinese Academy of Sciences (YZ201502), Youth Innovation Promotion Association CAS (2017175). (L. Meng, D. Dong, and X. Chen contributed equally to this work.) (Corresponding author: Z. Liu, J. Tian).

L. Meng, D. Dong, and M. Fang are with School of Artificial Intelligence, University of Chinese Academy of Sciences, Beijing, China, and with Key Laboratory of Molecular Imaging, Institute of Automation, Chinese Academy of Sciences, Beijing, China. X. Chen is with Department of Radiology, Guangzhou First People's Hospital, School of Medicine, South China University of Technology, Guangzhou, China. R. Wang is with Department of Radiology, Guizhou Provincial People's Hospital, Guiyang, China. J. Li is with Department of Radiology, Henan Cancer Hospital, Zhengzhou, China. Z. Liu is with Department of Radiology, Guangdong Provincial People's Hospital, Guangdong Academy of Medical Sciences, Guangzhou, 510080, China. (email: zyliu@163.com). J. Tian is with Key Laboratory of Molecular Imaging, Institute of Automation, Chinese Academy of Sciences, Beijing 100190, China, and with Beijing Advanced Innovation Center for Big Data-Based Precision Medicine, Beihang University, Beijing, 100191, China. (email: tian@ieee.org).

there is an inevitable dilemma when extracting features. Tumors are expressed in multiple layers in CT images, thus the features can be calculated with all involved layers (3D features) or just the single transverse layer that covers the largest area of the lesion (2D features). 2D annotations are easily performed with less labor consumption and lower computation complexity to calculate features, whereas 3D features might carry more tumor information. Therefore, there remains a long-time divergence about whether to use 2D or 3D annotations in GC radiomics-based research. To the best of our knowledge, the differences in characterizing GC between 2D and 3D features have not been reported yet.

In this study, we compared 2D and 3D radiomic features' representation and discrimination capability regarding three GC-related tasks: $T^{LNM}$, the prediction of lymph node metastasis (LNM, identified as pN0 stage in TNM staging system); $T^{LVI}$, the prediction of lymphovascular invasion (LVI); $T^{pT}$, the binary classification of whether pT4 or other pT stages. Based on four-center GC patients, we constructed six models ( $Model_{2D}^{LNM}$ , $Model_{3D}^{LNM}$ ; $Model_{2D}^{LVI}$ , $Model_{3D}^{LVI}$ ; $Model_{2D}^{pT}$ , $Model_{3D}^{pT}$ ) according to the collaboration of two modalities (2D, 3D) and three tasks. The performances of these models implied the GC-focused representation capability of the ROI and corresponding features. Besides, we designed an auxiliary experiment to explore the impact of resampling spacing settings, further to amplify our conclusion. Through this study, we hope to provide a related reference to further radiomics-based GC clinical researches.

Note that we use the word "modality" to express "2D" or "3D" in the paper.

## II. MATERIALS AND METHODS

### A. Patients

This multi-center retrospective study has attained ethical approval from the institutional review board in all participating four centers, and waived the informed consent requirement. *Supplementary Section A* shows the inclusion and exclusion criteria of the patient recruitment processes.

We enrolled 539 patients who were histologically confirmed GC and in the absence of preoperative therapy (radiotherapy, chemotherapy, or chemoradiotherapy). Among involved patients, 252 of them were from Guangdong Provincial People's Hospital (GDPPH), 181 from Guizhou Provincial People's Hospital (GZPPH), 82 from Henan Cancer Hospital (HCH), and 30 from Lanzhou University Second Hospital (LUSH). We integrated the data into a whole dataset and divided it into a training cohort ($n = 377$) and a validation cohort ($n = 162$) at a ratio of 7:3. The patients were partitioned by a computerized random number generator rather than by different source centers, which aimed to avoid that inter-center difference's impacts submerge 2D / 3D modality's nature.

The baseline demographic and clinicopathological characteristics were retrieved and shown in Table I. The detailed characteristics' distribution for each center are shown in *Supplementary Section B*. The pathology TNM staging of patients was confirmed according to the 8th American Joint

TABLE I
CHARACTERISTICS OF PATIENTS

| Characteristics | Training cohort | Validation cohort | P-value |
|---|---|---|---|
| Center, No. (%) | | | 0.9895 |
| GDPPH | 171 (45.3%) | 76 (46.9%) | |
| GZPPH | 128 (34.0%) | 53 (32.7%) | |
| HCH | 57 (15.1%) | 24 (14.8%) | |
| LUSH | 21 (5.6%) | 9 (5.6%) | |
| Sex, No. (%) | | | 0.5843 |
| Male | 257 (68.2%) | 115 (71.0%) | |
| Female | 120 (31.8%) | 47 (29.0%) | |
| Age, mean ± SD, years | 58.9 ± 11.5 | 59.2 ± 12.0 | 0.2232 |
| Location, No. (%) | | | 0.6802 |
| Proximal | 103 (27.3%) | 50 (30.9%) | |
| Middle | 77 (20.4%) | 33 (20.4%) | |
| Distal | 197 (52.3%) | 79 (48.7%) | |
| Differentiation, No. (%) | | | 0.7663 |
| Moderately or well differentiated | 126 (33.4%) | 57 (35.2%) | |
| Poorly differentiated or undifferentiated | 251 (66.6%) | 105 (64.8%) | |
| Lymph node metastasis [a], No. (%) | | | 0.3173 |
| Yes | 91 (24.1%) | 31 (19.1%) | |
| No | 286 (75.9%) | 131 (80.9%) | |
| Lymphovascular invasion [a], No. (%) | | | 0.8261 |
| Yes | 208 (55.2%) | 87 (53.7%) | |
| No | 169 (44.8%) | 75 (46.3%) | |
| pT stage, No. (%) | | | 0.9429 |
| I | 10 (2.6%) | 3 (1.9%) | |
| II | 27 (7.2%) | 11 (6.8%) | |
| III | 171 (45.4%) | 76 (46.9%) | |
| IV [a] | 169 (44.8%) | 72 (44.4%) | |

The distribution of patients' characteristics did not show significant differences (P-value > 0.05) between the training (n = 377) and validation (n = 162) cohorts. The P-value of "Age" was derived using Mann-Whitney U test, while other P-values were derived using $\chi^2$ test.
[a] The predicted labels: LNM, LVI, and pT4 stage.

Committee on Cancer (AJCC) 's suggestion [21]. LVI, defined as the presence of tumor emboli within either lymphatic or vascular, was ascertained with expertise in gastric histopathologic. Vascular channels were not differentiated from lymphatic vessels because of the difficulty and unsatisfactory reproducibility in clinical practice.

### B. CT Imaging Protocol and 2D/3D ROI Acquisition

Enrolled patients of the four centers underwent similar CT examinations but with different systems and protocols, which was clarified in *Supplementary Section C*. Venous phase CT images were retrieved from picture archiving and communication system.

Of each center, one or two radiologists with at least ten years of experience have reviewed all slices (transverse planes) of a patient's CT image and annotated tumor areas. A 3D ROI of the tumor was then segmented on the whole volume. For 2D ROI, we succinctly selected the slice with the largest tumor area annotated derived from the corresponding 3D ROI using an automatic algorithm.

After three months, 60 patients in the training cohort were randomly selected and performed ROIs segmentation again. The re-segmented ROIs were used to assess the reproducibility and stability of radiomic features by an interclass consistency test.

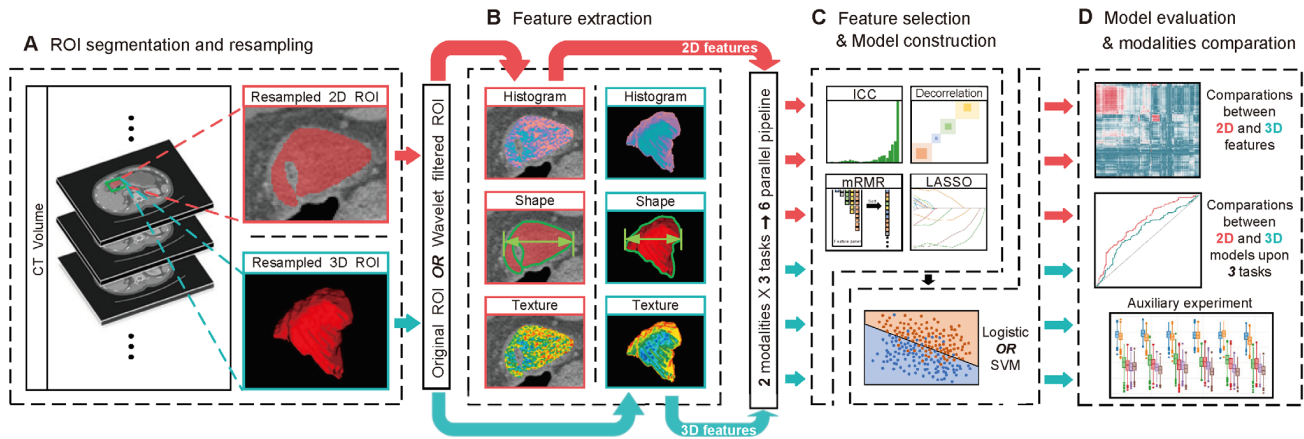

Fig. 1.  Radiomics processing flowchart. The red pipelines represent 2D modality part, while the cyan pipelines represent 3D. From the original CT volume, we obtained 2D and 3D ROIs (A), from which the radiomic features were extracted (B). After feature selection, three 2D-features-based models and three 3D-features-based were constructed corresponding three tasks (C). For each task, 2D- and 3D- features-based model were evaluated and compared (D). The implement of our workflow was based on Python (version 3.7.3; Wilmington, DE, USA; https://www.python.org/).

### C.  CT Image Resampling

Image resampling is a crucial but inexplicit preprocessing for radiomics-based researches, which were shown in the flow chart (Fig. 1, A).

On the one hand, for a single image volume, the CT image volumes usually have a smaller voxel spacing in the $x$- and y-axis than in the $z$-axis. It means that the spacing between slices (i.e. thickness) is larger compared to the in-plane spacing. To correctly extract and represent the spatial information of images, we need to resample the volume to guarantee the voxels' isotropic in three axes $(x, y, z)$. As to the resampled voxel spacing, we need to compromise between losing information in-plane (down-sampling) and interpolating information out-of-plane (up-sampling) [22], both of which will cause information distortion.

On the other hand, for different images, the reconstruction slice thickness varied from different CT protocols, especially for multi-center studies. To ensure the comparability of radiomic features between different patients, we need to resample all enrolled CT image volumes to a same isotropic voxel spacing. Therefore, we also need to compromise between losing high-resolution images' information and introducing artificial information to low-resolution images.

Obviously, $feature_{3D}$'s calculation demands more deliberate resampling options other than $feature_{2D}$'s. The former is synthesized from 3D ROIs with multi slices, whereas the latter is succinctly derived from single slice neglecting the thickness.

Our enrolled CT volumes were reconstructed with slice thicknesses of 1.00-5.00 mm (mean ± standard deviation, 3.11 ± 1.87 mm), and most of them were 1.25 mm or 5.00 mm. Considering both the distribution of our enrolled materials and the information fidelity, we resampled the 2D ROIs to a 1.25 mm × 1.25 mm voxel spacing to maintain the detailed in-plane information, and the 3D ROIs to 2.50 mm × 2.50 mm × 2.50 mm to compromise between in-plane information loss and out-of-plane information interpolation. We used different resampling settings to maximize the respective benefits of 2D and 3D modalities. The resampling algorithm was based on the B-spline interpolation [23]. Then, the resampled ROIs were fed into the processing pipelines parallelly.

In an auxiliary experiment, we have explored how the resampled voxel spacings impact the models' results, which will be detailed later.

### D.  Radiomic Feature Extraction

We extracted $features_{2D}$ and $features_{3D}$ respectively from 2D or 3D ROIs. The calculation of $features_{3D}$ was based on the whole 3D ROI instead of merging across slices. The features are defined in compliance with the Imaging Biomarker Standardization Initiative (IBSI) [22], according to the widely accepted studies [24], [25]. They were calculated on either original image, or a derived image obtained by applying one of the eight wavelet filters (eight combinations of applying either a high- or low-pass decomposition filter in three axes). The wavelet $feature_{2D}$'s calculation implicitly involves 3D spatial information because wavelet filters were applied on images' three axes. To guarantee the comparability between the two modalities, $features_{2D}$ and $features_{3D}$ were designed to be one-to-one correspondence and have the same amount. Both $features_{2D}$ and $features_{3D}$ consist of nine groups (one original feature group and eight wavelet feature groups). The original feature group was comprised of 7 feature classes, including 14 shape-based features, 18 first-order statistics features, 24 Gray Level Co-occurrence Matrix (GLCM) features, 16 Gray Level Run Length Matrix (GLRLM) features, 16 Gray Level Size Zone Matrix (GLSZM) features, 14 Gray Level Dependence Matrix (GLDM) features, and 5 Neighboring Gray Tone Difference Matrix (NGTDM) features.

Each wavelet feature group was comprised of all above except for shape-based features, which are independent of gray value and were extracted from the ROI masks. GLCM, GLRLM, GLSZM, GLDM, and NGTDM features are descriptors of image textures.

As a result, a total of 867 features for each modality, quantitatively representing phenotypic characteristics and heterogeneity of ROIs regarding GC, were extracted for each patient. Before further processing, we standardized each feature value to its z-score along with the patients using the trained parameters. The z-score of the feature $x$ of a patient is:

$$z = \frac{x - \mu_x}{\sigma_x} \qquad (1)$$

where $\mu_x$ and $\sigma_x$ are the mean and standard deviation of the population, respectively. The standardization can accelerate the following training process and enhance its stability. We used unsupervised clustering and a radiomic heatmap to explore the correlation between 2D and 3D feature bunches.

### E. Feature Selection

The feature selection strategy was subject to the features' stability and the discrimination performance. Firstly, to assess the reproducibility and robustness of radiomic features in the presence segmentation bias, we performed an interclass consistency test using the 60 re-segmented training images. Only the features with interclass consistency coefficient (ICC) > 0.75 were considered reproducible and selected.

Further, we fed $features_{2D}$ and $features_{3D}$ into the processing pipelines parallelly regarding three tasks, as shown in Fig. 1. The pipelines included the procedures as below:

#### 1) Distribution analysis

Mann–Whitney U test was used to measure the difference of each feature's distribution within the positive and negative sample groups.

$$U_i = R_i - \frac{n_i(n_i + 1)}{2} \qquad (2)$$

where $i$ is the positive or negative group, $n_i$ is the data size of the group $i$, and $R_i$ is the sum of the ranks in the group $i$. The smaller $U_i$ will be used to consult significant tables and derived a P-value. A smaller P-value indicates that the corresponding feature can distinguish positive and negative samples. We weeded out the features with a P-value higher than 0.05.

#### 2) Decorrelation

We calculated a Pearson linear correlation coefficient for every two features:

$$r = \frac{\sum_{n=1}^{N} \left( x_n^i - \overline{x^i} \right) \left( x_n^j - \overline{x^j} \right)}{\sqrt{\sum_{n=1}^{N} \left( x_n^i - \overline{x^i} \right)^2} \sqrt{\sum_{i=1}^{N} \left( x_n^j - \overline{x^j} \right)^2}} \qquad (3)$$

where $x^i$ and $x^j$ are two different features of the patients in the training cohort, $N$ is the data size of training cohort. For those feature pairs with a correlation coefficient surpassed 0.95, we only kept the one with a smaller P-value derived from the previous step.

#### 3) Minimum Redundancy Maximum Relevance (mRMR)

With the mRMR method [26], we selected the features which were mutually far away from each other while still highly correlated to the prediction label. The method is based on mutual information which is defined as:

$$I(x; y) = \iint p(x, y) \log \frac{p(x, y)}{p(x)p(y)} \qquad (4)$$

Assume that the we have $X$ features in total, and we have selected $m - 1$ features which formed the feature set $S_{m-1}$. We can select the $m$-th feature by incrementally optimizing the objective function:

$$\max_{x^j \in X - S_{m-1}} \left[ I(x^j; y) - \frac{1}{m-1} \sum_{x^i \in S_{m-1}} I(x^j; x^i) \right] \qquad (5)$$

where $y$ is the classification variables of the samples in the training cohort, $x^i$ and $x^j$ are different features of the patients in the training cohort.

#### 4) Least Absolute Shrinkage and Selection Operator (LASSO)

We selected the non-zero coefficient features as the final potential descriptor group for each specific task with LASSO, which is a linear model with an added $L_1$-norm regularization urging sparse variable coefficients [27]. The optimization objective for LASSO is:

$$\min_{\boldsymbol{\omega}} \sum_{n=1}^{N} (y_n - \boldsymbol{\omega}^T \boldsymbol{x}_n)^2 + \lambda \|\boldsymbol{\omega}\|_1 \qquad (6)$$

where $\boldsymbol{x}_n$ is the $n$-th patient's feature vector, $y_n$ is the classification variable, $\boldsymbol{\omega}$ is the weight vector of the linear model, and $\lambda > 0$ is the normalization parameter.

After going through the above procedures serially, the eligible features were finally selected.

All six processing pipelines, corresponding to three tasks $\times$ two models, shared the same structure as shown in Fig. 1 to guarantee the comparability. Additionally, we customized the specific details and parameters of each pipeline to drive the modalities to its optimal performances corresponding to every task.

### F. Models' Construction and Evaluation

After feature selection, we constructed and trained a multivariate prediction model respectively for each modality per task, leveraging the corresponding feature group. The five-fold cross-validation grid search method was used on the training cohort for model selection and parameter tuning. According to the performance, a logistic regression model or a support vector machine (SVM) with fine-tuned paraments was finalized for each of the six contexts (two modalities $\times$ three tasks). After that, we derived six models ($Model_{2D}^{LNM}$, $Model_{3D}^{LNM}$; $Model_{2D}^{LVI}$, $Model_{3D}^{LVI}$; $Model_{2D}^{pT}$, $Model_{3D}^{pT}$).

For each task ($T^{LNM}$, $T^{LVI}$, or $T^{pT}$), we evaluated and compared the corresponding $Model_{2D}$ and $Model_{3D}$'s performance, which reflected GC-focused representation and discrimination capacity of 2D or 3D modality. We used the receiver operating characteristic (ROC) curve, and area under the curve (AUC) [28] to assess the prediction performance. A higher AUC value indicates that the features of the corresponding modality can better represent and describe the characteristics of GC, and has better discrimination capacity for the prediction label. Considering the stability and generalization of the model among different data sources, 1000-time bootstrap was adopted among the training and validation cohorts, to estimate the AUC's variance and 95% CI.

Observations of this experiment would indicate that whether it is better to use 2D or 3D annotations in GC radiomics-based research. Based on the observations, we would further discuss the trade-offs between the choice of 2D or 3D in *Section IV*.

### G. Auxiliary Experiment

The above-mentioned main experiment was designed based on the fixed training and validation cohorts with fixed resampling settings (isotropic 1.25 mm for 2D ROIs, and 2.50 mm for 3D ROIs). Clinicians may be interested in whether the repartition of the cohorts, as well as the different resampling settings, would make the results any inconsistent with the previous observations. Hence, we designed an auxiliary experiment, verifying the validity and robustness of the observations, in order to strengthen our conclusion.

Specifically, we acquired five resampled ROIs groups leveraging five resampling spacing settings (triaxial isotropic 1.25 mm, 2.00 mm, 2.50 mm, 3.00 mm and 5.00 mm for both 2D and 3D modalities. Then we randomly repartitioned each ROI group into the training and validation cohorts (7:3), for 1000 times. Of each time, we fed the resampled and repartitioned 2D and 3D ROIs cohorts into the processing pipelines (Fig. 1) without fine-tuning. We forced the final selected features less than five, and then constructed six logistic regression models to obtain six AUCs (two modalities × three tasks). Finally, under every combined condition of five resampling settings, two modalities, and three tasks, we captured a total of 30 AUC distributions among the 1000-time cohort repartitions. We depicted three violin plots to visualize the AUC distributions of each task.

Moreover, clinicians may wonder, whether the combination of $features_{2D}$ and 3D spatial information would come to a better performance than $features_{2D}$ alone. Along with the auxiliary experiment, we expand the process flow (Fig. 1) with another "2.5D" pipeline, in which we synthesized the 2.5D features ($features_{2.5D}$) by averaging the $features_{2D}$ extracted from every slice of the 3D annotation. Note that the calculation of $features_{2.5D}$ is far more costly than $features_{2D}$ and $features_{3D}$. Other processes, such as model construction and evaluation processes, were parallel with 2D and 3D pipelines.

By comparing the distributions of 2D and 3D AUCs of every task, we could explore the impact of resampling settings and cohort partitions, further verify and amplify the observations of the main experiment and strengthen our conclusions.

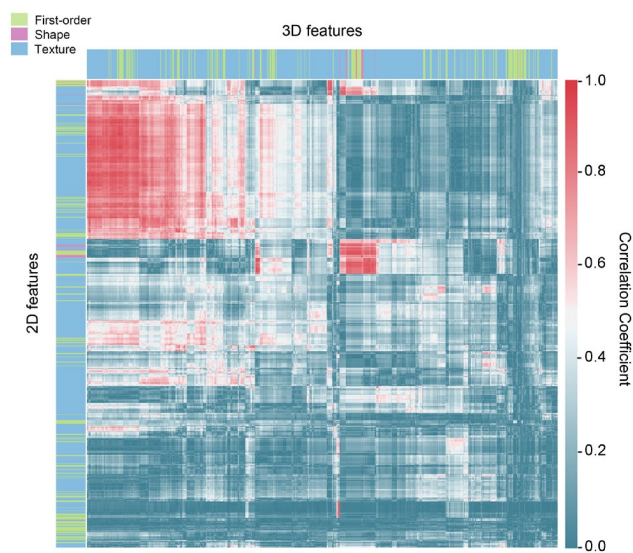

Fig. 2. Radiomic feature cluster heatmap. Redder regions imply that the corresponding 2D and 3D features has stronger correlations, while the regions that more cyan are on the contrary.

## III. RESULTS

The section *A-C* are related to the main experiment's result, and the section *D* describes the auxiliary experiment's result.

### A. Clinic Characteristics of Patients

In the main experiment, a total of 539 patients were enrolled in this retrospective study and divided into a training cohort ($n = 377$) and a validation cohort ($n = 162$). We analyzed the distribution differences of the clinic characteristics between the two cohorts, via $\chi^2$ test for categorical variables and Mann-Whitney U test for age. There were no statistical significances (two-sides, P-value > 0.05) found, and the detailed information are shown in Table I.

### B. Radiomic Feature Selection

From 2D and 3D resampled ROIs, we respectively extracted 867 radiomic features embracing the guideline of IBSI [22]. We analyzed the correlation between $features_{2D}$ and $features_{3D}$ group using unsupervised clustering and visualized it via a heatmap shown in Fig. 2. Most of the $feature_{2D}$-$feature_{3D}$ pairs showed a weak or no correlation, whereas about only 9% (65,885 in 867 × 867 pairs) of them showed stronger correlations (|correlation coefficient| > 0.6). Based on the interclass consistency test, 691 $features_{2D}$ and 786 $features_{3D}$ with ICC > 0.75 were considered robust and selected respectively.

Then the data parallely flowed through the processing pipelines. Consequently, we selected 7 $features_{2D}$ and 6 $features_{3D}$ for the task $T^{LNM}$; 3 $features_{2D}$ and 2 $features_{3D}$ for $T^{LVI}$; 7 $features_{2D}$ and 5 $features_{3D}$ for $T^{pT}$. The specific information of the selected features of each pipeline were listed in *Supplementary Section D*.

### C. Models' Construction and Evaluation

With respect to each of the three tasks, we constructed two models based on the selected $features_{2D}$ and $features_{3D}$ respectively. After fine-tuned, four ($Model_{2D}^{LNM}$, $Model_{3D}^{LNM}$, $Model_{2D}^{pT}$, $Model_{3D}^{pT}$) of them were logistic regression models, while the other two ($Model_{2D}^{LVI}$ and $Model_{3D}^{LVI}$) were support vector machines (SVM) using radial basis function (RBF) kernel [29]. The ROCs of the models were shown in Fig. 3, and the yielded AUCs were listed in Table II. The results showed $Models_{2D}$ revealed competitive performance compared with $Models_{3D}$ in both cohorts on every task. It suggested that there is no 3D radiomics calculation that is beneficial, and $features_{2D}$ can be a more potential descriptor and predictor than $features_{3D}$ on our GC radiomics-based research. To further prove this, we combined the selected 2D and 3D features, and constructed $Models_{2D+3D}$ for each task. The models showed similar performances to $Model_{2D}$, which also supported our conclusion. These were detailed in *Supplementary Section E*.

### D. Auxiliary Experiment

To explore the impact brought by different cohort partitions and resampling settings, we constructed and evaluated a total of 30,000 models corresponding to the collaborations of 1,000-time cohort partitions, five resampling settings, two modalities, and three tasks. The distributions of the AUCs were shown in Fig. 4, which have three insets corresponding to three tasks. Each inset delineated five distributions, each of which were derived from 1,000-time replicate experiments and corresponding to one resampling setting. The results revealed that 2D modality's performances generally surpasses 3D's with statistical significance (Mann-Whitney U test, P-value < 0.05) no matter the tasks and the resampling spacings, except for one situation. In this situation of task $T^{pT}$ with the ROIs resampled to 3 mm voxel spacing, 3D modality's performances slightly surpassed 2D's, whereas not statistically significant (P-value = 0.0746).

The average of AUC values and confidence intervals for each situation were listed in the Table III. Without fine-tuning, the average AUCs were lower than those in the main experiment. Regarding the 2.5D modality, we found that it had comparable performance with 2D modality in tasks $T^{LVI}$ and $T^{pT}$, whereas it requires extremely expensive calculation. It took an average of 0.28 seconds to calculate an ROI's 2D features, 1.00 seconds for 3D features, and 6.30 seconds for 2.5D features on our dataset. We need to emphasize that the 2.5D modality is not our primary concern, therefore the results of 2.5D modality were not depicted in Fig. 4 to avoid confusion.

### IV. DISCUSSION

In this study, we enrolled four-center GC patients and designed two experiments. Through the experiments, we exhaustively assessed 2D and 3D radiomic features' representation and discrimination capacity in contexts of three GC-related tasks ($T^{TNM}$, $T^{LVI}$, $T^{pT}$). We concluded that $features_{2D}$ showed comparable performance with $features_{3D}$ on these tasks in our multi-center dataset, which indicated that 2D annotations were recommended preferential. Through this study, we hope to provide a reference to further GC radiomics-based researches.

Whether to use 2D or 3D annotations remains a long-time debate in GC radiomics-based researches. Typically, radiomic features were extracted from ROIs, which need much expertise and human-effort manual annotations [7]. This laborious work occupies lots of human efforts, meanwhile the features' calculation needs costly computation. These problems, especially apparent in 3D modality, significantly restricts the development of related fields. In existing radiomics-based GC research, 3D annotations intuitively contain more information, which seems like the first choice [17], [18], [30]. Whereas, as an alternative trade-off, time-saving 2D annotations were also employed in some studies [19], [20], [31], [32]. Interestingly, our findings suggested that 2D annotations might be a preferred choice in GC radiomics-based researches because of the better performance, which seems quite counterintuitive, since 3D

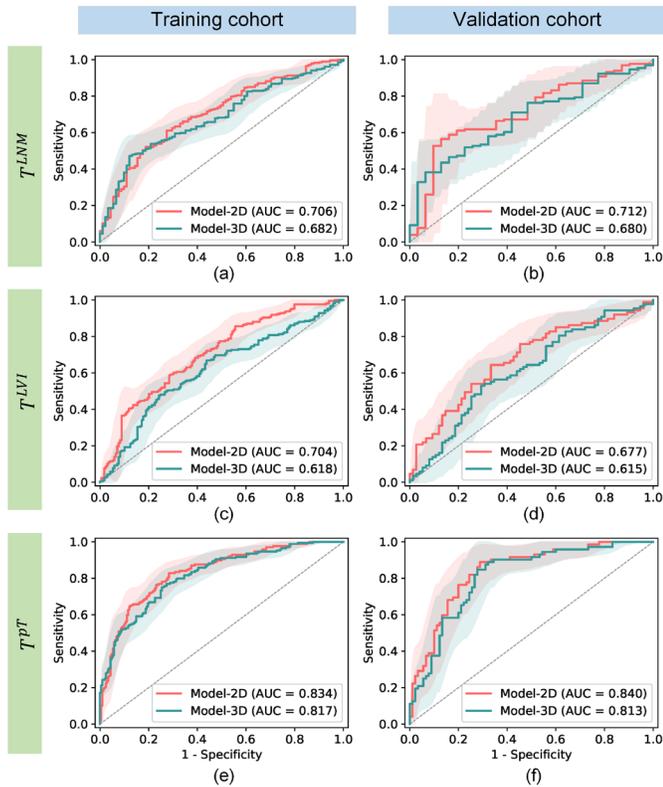

Fig. 3. Receiver operating characteristic (ROC) curves of three tasks in training and validation cohort. The shaded areas represent the 95% confidence interval. On every subgraph, 2D model has both an upper and lower confidence bound curve higher than 3D model.

TABLE II
THE PERFORMANCES OF THE MODELS

| Task | Model | Training cohort AUC [95% CI] | Validation cohort AUC [95% CI] |
|---|---|---|---|
| $T^{LNM}$ | $Model_{2D}^{LNM}$ | 0.706 [0.647, 0.765] | 0.712 [0.613, 0.811] |
|  | $Model_{3D}^{LNM}$ | 0.682 [0.624, 0.741] | 0.680 [0.584, 0.774] |
| $T^{LVI}$ | $Model_{2D}^{LVI}$ | 0.704 [0.652, 0.757] | 0.677 [0.595, 0.761] |
|  | $Model_{3D}^{LVI}$ | 0.618 [0.561, 0.674] | 0.615 [0.528, 0.703] |
| $T^{pT}$ | $Model_{2D}^{pT}$ | 0.834 [0.793, 0.875] | 0.840 [0.779, 0.901] |
|  | $Model_{3D}^{pT}$ | 0.817 [0.775, 0.859] | 0.813 [0.747, 0.879] |

The performances of the models corresponding different tasks and modalities. $Models_{2D}$ showed comparable performances with $Models_{3D}$ in both cohorts on all tasks.

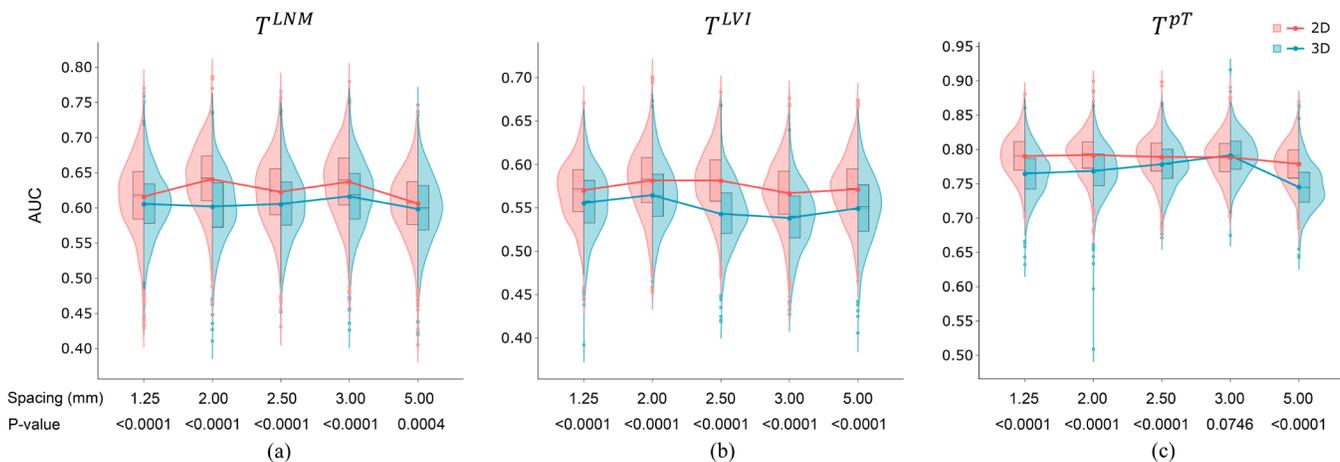

Fig. 4. The AUC distributions derived from 1,000-time replicate experiment. The P-values reflected the distribution differences of the 2D and 3D AUCs, and were derived by Mann-Whitney U test. Apart from the situation of task $T^{pT}$ with the 3.00 mm resampling spacing, other situations showed statistical significances (P-value < 0.05) between the AUC distributions of 2D (red) and 3D (cyan).

TABLE III
AUXILIARY EXPERIMENT'S RESULTS

| Task | Spacing (mm) | Average AUC on the validation cohort [95% CI] | | |
|---|---|---|---|---|
| | | 2D | 2.5D | 3D |
| $T^{LNM}$ | 1.25 | **0.615** **[0.728, 0.849]** | 0.608 [0.719, 0.838] | 0.605 [0.698, 0.831] |
| | 2.00 | **0.640** **[0.731, 0.848]** | 0.598 [0.732, 0.847] | 0.601 [0.722, 0.839] |
| | 2.50 | **0.622** **[0.509, 0.695]** | 0.591 [0.514, 0.696] | 0.605 [0.522, 0.710] |
| | 3.00 | **0.636** **[0.502, 0.638]** | 0.607 [0.506, 0.657] | 0.616 [0.511, 0.651] |
| | 5.00 | 0.606 [0.472, 0.626] | **0.620** **[0.501, 0.644]** | 0.598 [0.505, 0.646] |
| $T^{LVI}$ | 1.25 | 0.569 [0.699, 0.838] | **0.572** **[0.715, 0.841]** | 0.555 [0.732, 0.850] |
| | 2.00 | **0.581** **[0.513, 0.718]** | 0.575 [0.548, 0.733] | 0.564 [0.527, 0.718] |
| | 2.50 | **0.581** **[0.505, 0.692]** | 0.570 [0.517, 0.701] | 0.542 [0.502, 0.695] |
| | 3.00 | 0.566 [0.493, 0.640] | **0.573** **[0.501, 0.641]** | 0.538 [0.483, 0.627] |
| | 5.00 | 0.571 [0.497, 0.644] | **0.572** **[0.500, 0.647]** | 0.549 [0.498, 0.648] |
| $T^{pT}$ | 1.25 | 0.790 [0.731, 0.849] | **0.792** **[0.733, 0.850]** | 0.764 [0.729, 0.849] |
| | 2.00 | **0.791** **[0.681, 0.809]** | 0.787 [0.732, 0.853] | 0.768 [0.728, 0.847] |
| | 2.50 | 0.788 [0.541, 0.732] | **0.789** **[0.514, 0.698]** | 0.778 [0.515, 0.696] |
| | 3.00 | 0.788 [0.494, 0.689] | 0.789 [0.508, 0.708] | **0.791** **[0.524, 0.717]** |
| | 5.00 | 0.778 [0.492, 0.637] | **0.780** **[0.473, 0.613]** | 0.745 [0.467, 0.609] |

Validation cohort's average AUCs and 95% confidence intervals. **Bold letters** indicate the best performance among the three modalities when task and resampling spacing were fixed.

modality should carry more information.

However, we argue that 3D annotations can bring more noise, which submerges the efficacious information and interferes with the results. The noises mainly stem from two sources. Firstly, the voxel-wise annotation is a subjective and subtle procedure, in which whether lesion exists and where lesion boundaries lie can be quite divergent between different radiologists. The 3D modality would suffer more from this factor compared with 2D, because of its multi-slice annotation has amplified the influence. Secondly, the noises are related to the thicknesses: for a single image, the voxel spacing in the z-axis (thickness) is usually larger than the x-axis and y-axis (transverse planes); for different images, the thicknesses vary from different CT scanners' reconstruction protocols. Due to their multiple vague lesion boundaries and the higher susceptibility to different thicknesses, multi-slice 3D ROIs might have lower signal to noise ratio (SNR) than single-slice 2D ROIs.

Resampling images to triaxial isotropic can guarantee the extracted features maintain physical significance and comparable between multi-center images, but the inherent noises cannot be eliminated. In our main experiment, the interclass consistency test, which is the first feature selection step, indicated that $features_{3D}$ group possessed higher stability and reproducibility (ICC > 0.75) than the $features_{2D}$ group. We speculate that $features_{3D}$ absorbed and averaged multi-slice noises during synthesizing, which led to a more concentrated distribution. As a result, more $features_{3D}$ seemed uniform and stable like "white noise." In the following selection process, these less diverse features were weeded out. Finally, $Models_{3D}$ had fewer effective features supported, whereas $Models_{2D}$ were constructed based on more features with higher discriminability.

One may wonder what would happen if the cohorts are repartitioned and different resampling spacings are employed. To address the doubts, we designed the auxiliary experiment with a total of 30,000 models and derived AUC distributions. The results (Fig. 4, Table III) revealed that $Models_{2D}$ were statistically advantageous than $Models_{3D}$ in almost all circumstances, which strengthened our previous conclusion. For single-slice 2D ROIs, a smaller resampling spacing was intuitively considered better than a larger one because of the neglecting of thickness. However, we noticed that the models' performances did not present regular and consistent trends in the three tasks as the resampling spacing changed. The reason might be that smaller voxels are more sensitive to noise [25]. Therefore, a larger resampling spacing is not necessarily annoying, which only means that the extracted features focus on the coarser textures. Accompanying the auxiliary experiment, we explored a "2.5D" modality by combining $features_{2D}$ with 3D spatial information, whereas the overall performance did not exceed 2D modality.

A study conducted by *Welch et al.* suggested that many radiomic features may be dominated by the volume of ROI rather than its textures [33]. This research inspired us to conduct an experiment to explore whether the ROI volumes or shapes confounded the feature selection and model construction procedures. The experiment was detailed in *Supplementary Section F*, and the results of it still support our conclusion.

To our best knowledge, two studies discussed 2D and 3D radiomic features in other cancer, but their conclusions are contradictory and not sound enough to generalize to GC. *Shen et al.* [34] stated that 2D CT radiomics features outperformed 3D in non-small cell lung cancer's prognosis. Although they came to a similar conclusion to ours, the experiment was relatively simple. They did not consider the resampling settings, and the dataset was fixed without repartition. *Ortiz-Ramón et al.* [35] claimed that 3D MRI radiomic features surpassed 2D when classifying brain metastases from lung cancer and melanoma. However, the dataset was relatively small (30 patients), and the features were not sufficiently extracted (43 features). Our study has addressed the above-mentioned limitations and focused on GC. Moreover, we modestly hope to inspire other cancer's radiomics-based research through our comprehensive experiments.

Our study has several limitations. In terms of clinical problems, the study was based on gastric cancer. In a further study, we want to explore if similar observations would appear on other kinds of cancer. In terms of methodology, the study was confined to hand-craft radiomics features. In the future, we will investigate how the annotations and preprocessing of the materials would impact the performance of other methods, such as deep learning. Furthermore, if the appropriate image compensation method was applied on different centers, the model performances may be further improved. Moreover, the experiments were implemented on only diagnosis (binary classification) tasks. We will investigate the performances of 2D and 3D ROIs under the circumstances prognosis problems.

## V. Conclusion

This study compared the representation capacity and discriminability of 2D and 3D radiomic features in gastric cancer through three clinic problems. Models constructed with 2D radiomic features revealed comparable performance in the characterization of preoperative GC compared with those constructed with 3D features. Therefore, one-slice 2D annotations are recommended to use preferentially considering the comparable performance of characterizing GC and less laborious manual annotation.

## References

[1] F. Bray, J. Ferlay, I. Soerjomataram, R. L. Siegel, L. A. Torre, and A. Jemal, "Global cancer statistics 2018: GLOBOCAN estimates of incidence and mortality worldwide for 36 cancers in 185 countries," *CA. Cancer J. Clin.*, vol. 68, no. 6, pp. 394–424, 2018, doi: 10.3322/caac.21492.

[2] M. Roberto *et al.*, "Prognosis of elderly gastric cancer patients after surgery: a nomogram to predict survival," *Med. Oncol.*, vol. 35, no. 7, p. 111, 2018.

[3] X.-C. Shang-Guan *et al.*, "Preoperative lymph node size is helpful to predict the prognosis of patients with stage III gastric cancer after radical resection," *Surg. Oncol.*, vol. 27, no. 1, pp. 54–60, 2018.

[4] T. Kinoshita and A. Kaito, "Current status and future perspectives of laparoscopic radical surgery for advanced gastric cancer," *Transl. Gastroenterol. Hepatol.*, vol. 2, 2017.

[5] J. A. Ajani *et al.*, "Gastric cancer, version 3.2016, NCCN clinical practice guidelines in oncology," *J. Natl. Compr. Cancer Netw.*, vol. 14, no. 10, pp. 1286–1312, 2016.

[6] R. H. de O. Barros, T. J. Penachim, D. L. Martins, N. A. Andreollo, and N. M. G. Caserta, "Multidetector computed tomography in the preoperative staging of gastric adenocarcinoma," *Radiol. Bras.*, vol. 48, no. 2, pp. 74–80, 2015.

[7] P. Lambin *et al.*, "Radiomics: extracting more information from medical images using advanced feature analysis," *Eur. J. Cancer*, vol. 48, no. 4, pp. 441–446, 2012.

[8] W. L. Bi *et al.*, "Artificial intelligence in cancer imaging: clinical challenges and applications," *CA. Cancer J. Clin.*, vol. 69, no. 2, pp. 127–157, 2019.

[9] D. Dong *et al.*, "Development and validation of a novel MR imaging predictor of response to induction chemotherapy in locoregionally advanced nasopharyngeal cancer: a randomized controlled trial substudy (NCT01245959)," *BMC Med.*, vol. 17, no. 1, p. 190, 2019.

[10] J. Song *et al.*, "A new approach to predict progression-free survival in stage IV EGFR-mutant NSCLC patients with EGFR-TKI therapy," *Clin. Cancer Res.*, vol. 24, no. 15, pp. 3583–3592, 2018.

[11] J. Li *et al.*, "Diagnostic accuracy of dual-energy CT-based nomograms to predict lymph node metastasis in gastric cancer," *Eur. Radiol.*, vol. 28, no. 12, pp. 5241–5249, 2018.

[12] Z. Liu *et al.*, "The applications of radiomics in precision diagnosis and treatment of oncology: Opportunities and challenges," *Theranostics*, vol. 9, no. 5, pp. 1303–1322, 2019, doi: 10.7150/thno.30309.

[13] Y. Zhou *et al.*, "A radiomics approach with CNN for shear-wave elastography breast tumor classification," *IEEE Trans. Biomed. Eng.*, vol. 65, no. 9, pp. 1935–1942, 2018, doi: 10.1109/TBME.2018.2844188.

[14] A. Cameron, F. Khalvati, M. A. Haider, and A. Wong, "MAPS: A Quantitative Radiomics Approach for Prostate Cancer Detection," *IEEE Trans. Biomed. Eng.*, vol. 63, no. 6, pp. 1145–1156, 2016, doi: 10.1109/TBME.2015.2485779.

[15] T. T. Zhai *et al.*, "Pre-treatment radiomic features predict individual lymph node failure for head and neck cancer patients," *Radiother. Oncol.*, vol. 146, pp. 58–65, 2020, doi: 10.1016/j.radonc.2020.02.005.

[16] Y. S. Choi *et al.*, "Machine learning and radiomic phenotyping of lower grade gliomas: improving survival prediction," *Eur. Radiol.*, 2020, doi: 10.1007/s00330-020-06737-5.

[17] Y. Wang *et al.*, "Prediction of the Depth of Tumor Invasion in Gastric Cancer: Potential Role of CT Radiomics," *Acad. Radiol.*, 2019.

[18] Z. Ma *et al.*, "CT-based radiomics signature for differentiating Borrmann type IV gastric cancer from primary gastric lymphoma," *Eur. J. Radiol.*, vol. 91, pp. 142–147, 2017.

[19] Y. Jiang *et al.*, "Radiomics Signature on Computed Tomography Imaging: Association With Lymph Node Metastasis in Patients With Gastric Cancer," *Front. Oncol.*, vol. 9, p. 340, 2019.

[20] Y. Jiang *et al.*, "Radiomics signature of computed tomography imaging for prediction of survival and chemotherapeutic benefits in gastric cancer," *EBioMedicine*, vol. 36, pp. 171–182, 2018.

[21] M. B. Amin *et al.*, "The Eighth Edition AJCC Cancer Staging Manual: Continuing to build a bridge from a population-based to a more 'personalized' approach to cancer staging," *CA. Cancer J. Clin.*, vol. 67, no. 2, pp. 93–99, 2017.

[22] A. Zwanenburg, S. Leger, M. Vallières, and S. Löck, "Image biomarker standardisation initiative," *CancerData*, Dec. 2016, doi: 10.17195/candat.2016.08.1.

[23] T. M. Lehmann, C. Gonner, and K. Spitzer, "Addendum: B-spline interpolation in medical image processing," *IEEE Trans. Med. Imaging*, vol. 20, no. 7, pp. 660–665, 2001.

[24] H. J. W. L. Aerts *et al.*, "Decoding tumour phenotype by noninvasive imaging using a quantitative radiomics approach," *Nat. Commun.*, vol. 5, 2014, doi: 10.1038/ncomms5006.

[25] J. J. M. Van Griethuysen *et al.*, "Computational radiomics system to decode the radiographic phenotype," *Cancer Res.*, vol. 77, no. 21, pp. e104–e107, 2017.

[26] H. Peng, F. Long, and C. Ding, "Feature selection based on mutual information: criteria of max-dependency, max-relevance, and min-redundancy," *IEEE Trans. Pattern Anal. Mach. Intell.*, no. 8, pp. 1226–1238, 2005.

[27] R. Tibshirani, "Regression shrinkage and selection via the lasso: a retrospective," *J. R. Stat. Soc. Ser. B (Statistical Methodol.*, vol. 73, no. 3, pp. 273–282, 2011.

[28] K. H. Zou, A. J O'Malley, and L. Mauri, "Receiver-operating characteristic analysis for evaluating diagnostic tests and predictive models," *Circulation*, vol. 115, no. 5, pp. 654–657, 2007.

[29] F. R. Bach, G. R. G. Lanckriet, and M. I. Jordan, "Multiple kernel learning, conic duality, and the SMO algorithm," in *Proceedings of the twenty-first international conference on Machine learning*, 2004, p. 6.

[30] Y. Wang *et al.*, "CT radiomics nomogram for the preoperative prediction of lymph node metastasis in gastric cancer," *Eur. Radiol.*, pp. 1–11, 2019.

[31] D. Dong *et al.*, "Development and validation of an individualized nomogram to identify occult peritoneal metastasis in patients with advanced gastric cancer," *Ann. Oncol.*, vol. 30, no. 3, pp. 431–438, 2019, doi: 10.1093/annonc/mdz001.

[32] D. Dong *et al.*, "Deep learning radiomic nomogram can predict the number of lymph node metastasis in locally advanced gastric cancer: an international multi-center study," *Ann. Oncol.*, 2020, doi: 10.1016/j.annonc.2020.04.003.

[33] M. L. Welch *et al.*, "Vulnerabilities of radiomic signature development: The need for safeguards," *Radiother. Oncol.*, vol. 130, pp. 2–9, 2019, doi: 10.1016/j.radonc.2018.10.027.

[34] C. Shen *et al.*, "2D and 3D CT Radiomics Features Prognostic Performance Comparison in Non-Small Cell Lung Cancer," *Transl. Oncol.*, vol. 10, no. 6, pp. 886–894, 2017, doi: 10.1016/j.tranon.2017.08.007.

[35] R. Ortiz-Ramon, A. Larroza, E. Arana, and D. Moratal, "A radiomics evaluation of 2D and 3D MRI texture features to classify brain metastases from lung cancer and melanoma," *Proc. Annu. Int. Conf. IEEE Eng. Med. Biol. Soc. EMBS*, no. July, pp. 493–496, 2017, doi: 10.1109/EMBC.2017.8036869.